\documentclass{aastex}
\usepackage{emulateapj5}
\usepackage{apjfonts}

\usepackage{natbib}

\shorttitle{Evolution of the Tully--Fisher relation}
\shortauthors{Ziegler et al.}

\begin{document}

\bibliographystyle{abbrvnat}


\title{The evolution of the Tully--Fisher relation of spiral
galaxies\altaffilmark{1}}
\altaffiltext{1}{Based on observations collected at the European
Southern Observatory, Cerro Paranal, Chile (ESO Nos. 65.0-0049 \& 66.A-0547).}


\author{B. L. Ziegler\altaffilmark{2}, A. B\"ohm, K. J. Fricke, K. J\"ager, 
H. Nicklas}
  \affil{Universit\"atssternwarte G\"ottingen, Geismarlandstr. 11, 37083 
  G\"ottingen, Germany}
\altaffiltext{2}{bziegler@uni--sw.gwdg.de}
\author{R. Bender, N. Drory, A. Gabasch, R. P. Saglia\altaffilmark{3}, 
S. Seitz}
  \affil{Universit\"atssternwarte M\"unchen, Scheinerstr. 1, 81679 M\"unchen, 
Germany}
\altaffiltext{3}{Present address: Research School of Astronomy and 
Astrophysics, The Australian National University, Cotter Road, Weston Creek, 
ACT 2611, Australia}
\and
\author{J. Heidt, D. Mehlert, C. M\"ollenhoff, S. Noll, 
E. Sutorius\altaffilmark{4}}
  \affil{Landessternwarte Heidelberg, K\"onigstuhl, 69117 Heidelberg, Germany}
\altaffiltext{4}{Present address: Royal Observatory Edinburgh, Blackford Hill,
Edinburgh EH9 3HJ, United Kingdom}


\begin{abstract}
We present the $B$-band Tully--Fisher relation (TFR) of 60 late--type galaxies
with redshifts $0.1-1$. The galaxies were selected from 
the FORS Deep Field with a limiting magnitude of $R=23$.
Spatially resolved rotation curves were derived from
spectra obtained with FORS2 at the VLT.
High-mass galaxies with $v_{\rm max}\gtrsim150$\,km/s show little evolution, 
whereas the least massive systems in our sample are brighter by $\sim1-2$ mag 
compared to their local counterparts. For the entire distant sample, 
the TFR slope is flatter
than for local field galaxies ($-5.77\pm0.45$ versus $-7.92\pm0.18$).
Thus, we find evidence for evolution of
the slope of the TFR with redshift on the $3\,\sigma$ level.
This is still true when we subdivide the sample into three redshift bins.
We speculate that the flatter tilt of our sample is caused by the evolution
of luminosities and an additional population of blue galaxies at $z\gtrsim0.2$.
The mass dependence of the TFR evolution also leads to variations for 
different galaxy types in magnitude-limited samples,
suggesting that selection effects can account for the discrepant
results of previous TFR studies on the luminosity evolution of late--type 
galaxies.
\end{abstract}




\section{Introduction}

\citet{TF77} found a remarkable correlation between the stellar
content and the kinematics of disk galaxies linking tightly the luminosity
to the maximum rotational velocity of late--type galaxies. 
This ``Tully--Fisher Relation'' (TFR) can therefore be exploited to study 
both the dynamical evolution and the star formation history of disk galaxies.
Observed local TFR samples \citep[e.g.][]{MF96} 
with their very small scatter already served as important 
constraints for galaxy evolution models \citep[e.g.][]{EL96}.
But the redshift evolution of the TFR is an even more powerful tool to
investigate galaxy evolution and cosmological structure formation.

On the observational side, recent studies of spiral galaxy kinematics at 
intermediate redshift produced rather discrepant results
mainly due to different sample selection methods. 
While \citet{VFPGFIK96} find only a modest increase in luminosity
of $\Delta M_B \approx 0.5$ at $\langle z \rangle \approx 0.5$,
the results of \citet{RGCI97} and \citet{SP98} imply a strong
brightening of roughly 2\,mag in absolute $B$-band luminosity.

Theoretical simulations have to consider the temporal growth of galaxies
in size and mass as well as the evolution of the star formation rates.
\citet{MMW98a}, for example, combine the dynamical evolution as traced by
Cold Dark Matter (CDM) hierarchical-merging models with a simple scaling 
relation between mass and luminosity.
\citet{SN99}, in addition, introduce star formation phenomena via
semianalytical recipes.
These kind of studies predict zeropoint offsets in luminosity for distant TFR 
samples depending both on observed wavelength and cosmology.

\section{Sample and Observations}

Our galaxy sample is based on the FORS Deep Field (FDF), a deep multi-color
study ($UBgRIzJK$) of an $\approx6'\times6'$ field with limiting blue and
visual magnitudes comparable to the HST Deep Fields \citep{ABBDF00,HABBD01}.
Photometric redshifts and spectrophotometric galaxy types
were determined for an $I$-band selected catalogue of $\approx3800$ objects
\citep{BABDF01}. From this list, late-type galaxies were drawn primarily 
on the basis of their brightness with a limiting magnitude of $R < 23^m$
\citep[as determined by the SExtractor package,][]{BA96}. 
Additional criteria have been
that the inclination was $i>40\degr$ and $z_{\rm phot}$ $<1.2$ so that any
[O\,{\sc ii}]3727 emission-line doublet would fall onto our observed
wavelength range. The final sample comprised 79 spirals of all Hubble types
(Sa -- Irr) with $0.1<z<1.0$ peaking at $z=0.35$ and $0.75$ (see
Fig.\,\ref{zhist}).
Their absolute $B$ magnitudes cover the range 
$-22.4 \le M_B \le -15.8$, the scale lengths range from 0.6\,kpc to 4.8\,kpc
($H_0=75$\,km\,s$^{-1}$Mpc$^{-1}$ and $q_0=0$ assumed throughout the paper).

\vbox{
\begin{center}
\leavevmode
\hbox{%
\epsfxsize=8cm
\epsffile{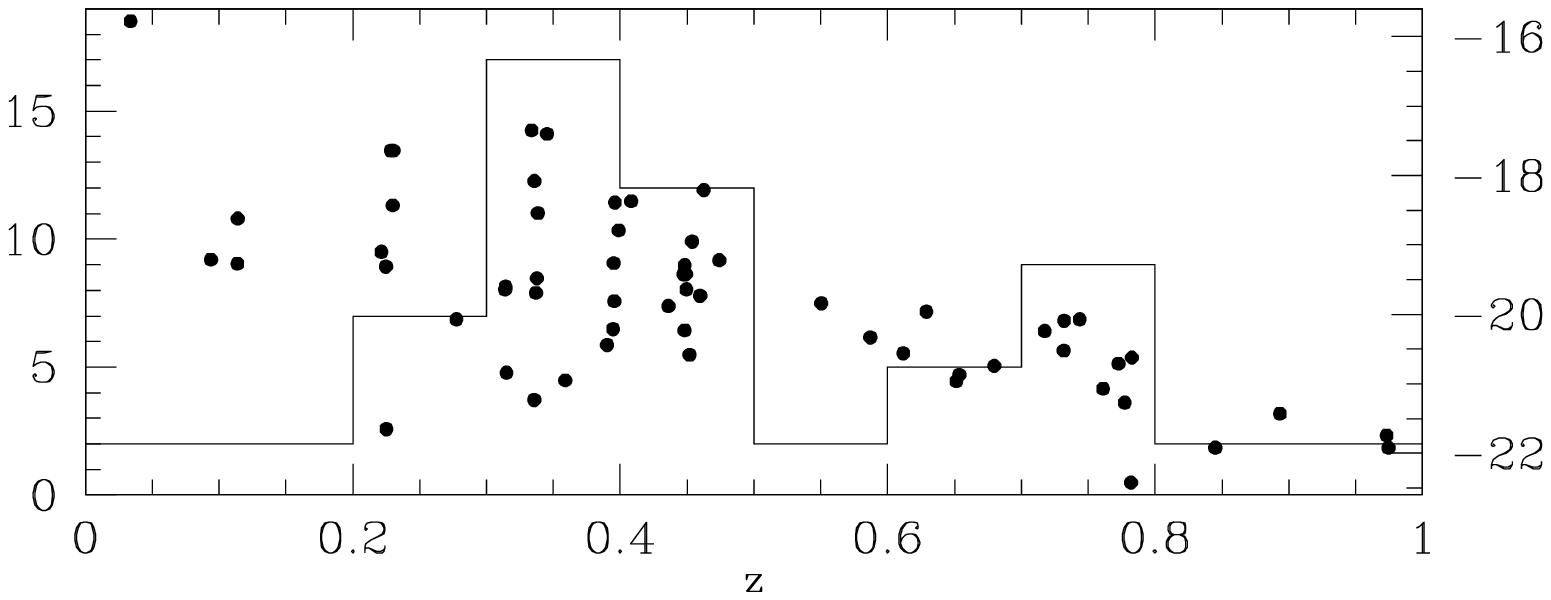}}
\figcaption{\small
\label{zhist}
Redshift histogram of the FDF sample. Overplotted is the $M_B$ distribution.}
\end{center}}


Spectroscopy was carried out with the Focal Reducer and Spectrograph 2 (FORS2)
mounted onto the Very Large Telescope (VLT) Unit 2. 
Using the grism \textsf{600R} with the order separation filter
\textsf{GG435}, the typical wavelength range for the case of a chip-centered 
slit was $\lambda\lambda=5250-7450$\,\AA. 
With a slit width of $1\arcsec$ the
dispersion was 1.08\,\AA/pixel and the resolution $R\approx1230$. Observations
were done in MOS (Multi Object Spectroscopy) mode, in which FORS2 provides
19 slitlets of $\approx22\arcsec$ length in a vertical array. 
Targets for the different setups were selected in a way to minimize the
misalignment between slit direction and position angle of the galaxy. The
maximal deviation was $15\degr$.
The exposures for each setup
were split into $3\times3000$\,s, the mean DIMM seeing was $0.66\arcsec$ 
(FWHM) for a spatial scale of $0.2\arcsec$/pixel.
A detailed description of the reduction process will be given in \citet{BZ02}.

\section{Determination of $v_{\rm max}$ and M$_{\rm B}$}

To derive the rotation velocities of the galaxies as a function of radius,
Gaussian profiles were fitted to the visible emission line(s). For the objects
with high redshift ($0.4\lesssim z\lesssim 1.0$), the [O\,{\sc ii}]3727 
doublet was used, while for objects with $0.1\lesssim z\lesssim 0.5$,
the [O\,{\sc iii}]5007 and/or H$_\beta$ lines were measured.
To increase the S/N, a boxcar filter of 0.6\arcsec\ was applied,
i.e. three neighbouring
rows of the spectra were averaged before each Gaussian fit.   

In contrast to studies of local spirals, these ``straightforward'' measurements
of $v_{\rm rot}$(r) cannot be used to directly derive the maximum rotation 
velocity $v_{\rm max}$ \citep{SP99}. 
Because the size of the visible disks of spirals at $z\approx0.5$ 
(with typical apparent scale lengths of r$_{\rm d}\approx0.5''$)
is comparable to the slit width of 1\arcsec, the spectroscopy is an
``integration'' over the galaxies' intrinsic velocity fields. 

To tackle this problem, 
a synthetic rotation curve (RC) was generated for each galaxy,
assuming a linearly rising rotation velocity in the
innermost galaxy regions and a constant rotation $v_{\rm max}$ at large radii
\citep[e.g.][]{Court97}.
The radius at which this intrinsic RC 
flattens was computed from the 
scale length $r_{\rm d}$ measured in the $I$-band, taking into account that
its value depends on the rest frame wavelength, because the scale length of
the young star population is larger than that of the older population 
\citep{RD94}. 
This intrinsic one-dimensional RC was then used to generate the
2D velocity field taking into account inclination and position angle.
The structural parameters were derived from two-dimensional luminosity 
profile fits based on a coadded I-band image with an FWHM of 0.5\arcsec. 
These fits considered the possible dependence of the Point Spread Function 
on CCD position. If a bulge was detectable, the fits included two components. 
To test our procedure, we performed a ``VLT simulation'' based on the drizzled 
HDF images by rebinning them to 0.2\arcsec/pixel and convolving them to 
0.5\arcsec\ FWHM.
The disc inclinations we measured in these frames were then compared with the 
values given by \citet{MS98} based on the original images. For 
objects with B/T<0.4, the differences were limited to a few degrees.
Estimated errors on the inclinations range from 7$\degr$ for
$i\approx40\degr$ to 3$\degr$ for $i\approx80\degr$, in some cases they may 
be as high as up to 12$\degr$.
In the next step, the inclined velocity field was
folded with the luminosity profile and convolved 
to account for the seeing during spectroscopy. 
A simulated RC was extracted from the 2D velocity field by integrating 
over the slit aperture.
At last, we variied $v_{\rm max}$, the only remaining free parameter in these 
simulations, until the best reproduction of the observed RC was achieved.
In that manner we were able to quantify the maximum rotation velocity 
of 60 spirals in our sample, with estimated errors between 10 and 40
km/s, depending on galaxy size and RC quality. The remaining 19 objects showed 
either a solid--body rotation, highly disturbed RCs or no rotation at all. 
Examples of RCs are given in Fig.\,\ref{rc}.

\vbox{
\begin{center}
\leavevmode
\hbox{%
\epsfxsize=7cm
\epsffile{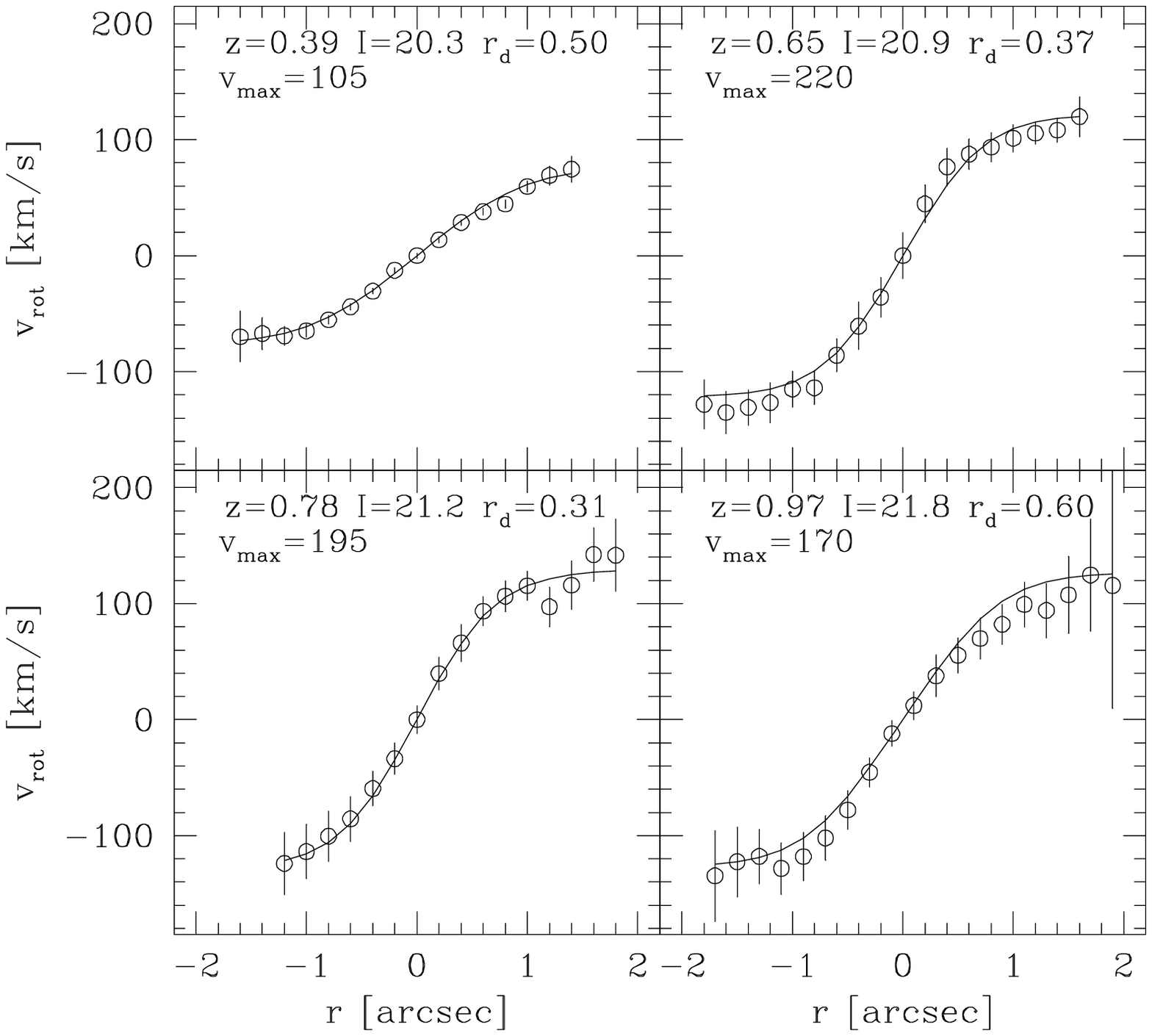}}
\figcaption{\small
Examples of observed rotation curves with their 1D--fit. 
Respective values of redshift, $I_{\rm obs}$ magnitude, disk scale length 
(in arcsec) and 
$v_{\rm max}$ (as derived with the 2D--model of the velocity field) are given. 
Errorbars denote the errors of the Gaussian fits to the line positions.
\label{rc}}
\end{center}}

The $B$-band luminosities of the FDF galaxies were determined in the following
way: we took SExtractor's {\sf MAG\_ISO} values as the total magnitudes in
the $BgRI$ FORS2 filters, which were then transformed onto the 
Johnson--Cousins photometric system. 
One origin of uncertainty in the derivation of the rest--frame $B$-band
luminosity is the $k$-correction \citep[see e.g.][]{ZSBBGS99}. 
For early-type spirals at $z\approx1.0$, 
transformation from comoving to rest--frame $B$-band magnitude
amounts to $\gtrsim2$\,mag.
To keep the $k$-corrections small, we therefore used that filter which best 
matches the rest--frame $B$.
Depending on the galaxies redshift, this was either the $B$-, $g$-, $R$- or
$I$-band, leading to a typical correction of 0.5\,mag. 
For the calculation of the respective $k$-corrections, we used template 
spectra created by the chemically consistent evolutionary synthesis
models of \citet{MFC01}. 
Observed galaxy colors of different Hubble types (e.g. \citet{FG94}) are
reproduced with these spectra to within 0.15\,mag over the whole covered
redshift range.
Corrections for inclination-dependent intrinsic absorption due to dust
were applied following \citet{TF85}.

\section{The Tully--Fisher relation at $z\approx0.5$}

For the comparison to local data, we chose the large sample of 1196 field
Sbc and Sc galaxies of \citet{HGCCF99}. 
To be able to better compare our results with previous studies, 
we construct the $B$-band TFR in this letter
\citep[for the $I$-band and NIR TFR see][]{BZ02}.
Therefore, the observed $I$ magnitudes given by Haynes et al. were
transformed to $B$ using the type-dependent ($B-I$) colors of \citet{FG94}
and corrected for intrinsic absorption the same way as  our sample.   
This transformation was tested on our sample (for which we both have
$B$ and $I$ magnitudes), showing that on average the accuracy is within 
0.1\,mag independent of galaxy type. The resulting absolute magnitudes cover
$-22.5 \le M_B \le -15.1$.
For use in the TFR, we further corrected the derived $M_B$ magnitudes
for morphological and incompleteness bias following \citet{GHHVC97}.
The authors pointed out that an incompleteness bias arises from
the differences between the luminosity function of a magnitude limited
sample and the Schechter luminosity function.
At low rotation velocities (low absolute magnitudes),
galaxies falling below the detection limit do not contribute to the
observed TFR, resulting in an underestimated slope. 
Simulating this, a synthetic, unbiased TFR is transformed into the observed 
relation using the measured values of $v_{\rm max}$, the observed scatter 
(as a function of $v_{\rm max}$) and the completeness limit.
The effect can then be counter-balanced by decreasing the observed luminosities
accordingly. 

The magnitudes of the distant galaxies were de-biased with the same procedure
adopting the parameterization of the scatter given by Giovanelli et al. but
taking into account that the scatter in the FDF sample is on the mean 1.5 
times larger. To account for the variation in the luminosity function with
redshift, we subdivided our full FDF sample into three $z$ bins,
for which we constructed the observed luminosity functions separately.
At the faint end of our sample, the correction for incompleteness amounts to
maximal $+1.4$\,mag.
The morphological bias, as a similar effect, originates in the
different luminosity functions for different morphological classes and leads
to a slightly increased luminosity for early types ($\sim-0.3...-0.1$\,mag).
If not accounted for, the combination of these two biases leads to
an underestimate of both slope and scatter of the TFR. 
We subdivided the 60 galaxies into 3 classes 
corresponding to Sa/Sb, Sc and Sd/Irr. This was done on the basis of 
spectrophotometric types from our photometric redshift catalog 
 and by comparing our spectra to the catalog of template spectra
of \citet{Kenni92}. There are 10 Sa/b, 32 Sc and 18 Sd/Irr galaxies in our
sample. 

We present the distant $B$-band TFR in Fig.\,\ref{tfb} in comparison to the 
local TFR. The FDF spirals follow a flatter relation than their local 
counterparts. While the most massive galaxies are indistinguishable on average
from the local sample, the less massive galaxies show large deviations from
the local fit. For the full FDF sample we find a slope of $-5.77\pm0.45$ by
repeating bisector fits \citep[e.g.][]{ZBSDL01} to the data centered on the 
median value of the $\log v_{\rm max}$-distribution 
($\langle\log v_{\rm max}\rangle = 2.08$) 100 times using a bootstrap method.
For the local sample ($\langle\log v_{\rm max}\rangle = 2.182$) the same 
method yields a slope of $-7.92\pm0.18$. The last value is consistent with
slopes derived for other local field samples. We derive, for example, a slope 
of $-7.40$ for the BBFN/RC3 sample \citep[not de-biased;][]{BBFN97}, 
and $-7.50$ for the \citet{PT92} sample.
\citet{SMHHM00} find a slope of $-8.07\pm0.72$ in the $B$-band using 
HST-based calibrators.
Thus, the slopes of the local samples and the FDF sample differ on the 
$3\,\sigma$ level. 
This is confirmed by a 2D Kolmogorov--Smirnov test, which limits the
probability that our sample originates from the same distribution as the
Haynes et al. sample to 0.001.
We also performed a simple Monte--Carlo simulation extracting randomly 60 
galaxies from the local sample. Neither of 200 iterations resulted in a TFR 
slope as flat as for the FDF galaxies.
If we subdivide our sample according to galaxy type (Sa/b, Sc and Sd/Irr)
or redshift, the resulting slopes and zero-points are compatible with each
other within their $1\sigma$ errors (see Table\,1).

\vbox{
\begin{center}
\leavevmode
\hbox{%
\epsfxsize=8cm
\epsffile{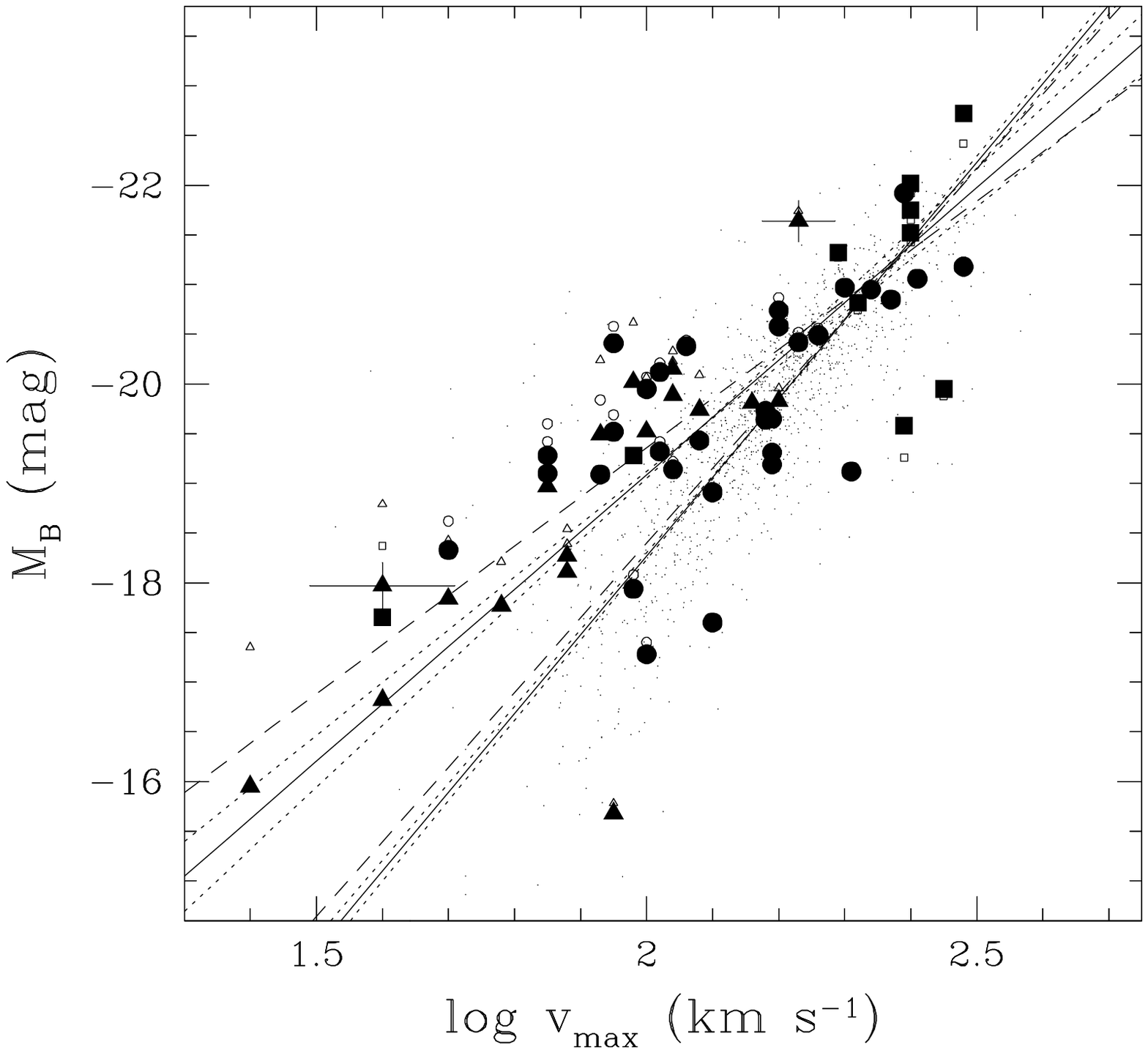}}
\figcaption{\small
The $B$-band Tully--Fisher relation of 60 field spiral
and irregular galaxies at intermediate redshifts. Big symbols denote our
spectrophotometric classification: squares Sa/Sb, circles Sc, triangles Sd/Irr.
Open symbols correspond to ``observed'' data, filled symbols to de-biased
data.
Solid (dotted) lines represent the best ($\pm1\sigma$) fits of the 100 
bootstrap bisector fits to the de-biased sample. The dashed lines are the 
fits to the ``observed'' data.
The steeper lines show similar 
fits to the large sample of 1200 local field spirals (dots) from 
\citet{HGCCF99}. 
There is a clear change of slope (on the $3\,\sigma$ level) between the local
and distant TFR.
Two typical errorbars are given.
\label{tfb}}
\end{center}}

%
%
\begin{center}  
{\scriptsize 
\centerline{\sc Table 1}
\centerline{\sc TFR slopes with $1\,\sigma$ errors}
\vspace{0.1cm}
\begin{tabular}{lccc} 
\hline\hline
\noalign{\smallskip}
sample & $N$ & $M_B$ vs. $\log v$ & $\log v$ vs. $M_B$ \\
\noalign{\smallskip}
\hline
\noalign{\smallskip}
distant sample: & & & \cr
FDF (not de-biased) & 60 & $-4.99\pm0.48$ & $-0.202\pm0.016$ \cr
FDF & 60 & $-5.77\pm0.45$ & $-0.175\pm0.014$ \cr
FDF $z<0.3$ & 11 & $-5.46\pm0.68$ & $-0.195\pm0.054$ \cr
FDF $0.3\le z<0.5$& 29 & $-4.62\pm0.48$ & $-0.220\pm0.026$ \cr
FDF $z\ge0.5$ & 20 & $-5.59\pm0.61$ & $-0.174\pm0.014$ \cr
\noalign{\smallskip}
\noalign{\hrule}
\noalign{\smallskip}
local samples: & & & \cr
Haynes et al. (not de-biased) & 1296 & $-7.52\pm0.18$ & $-0.133\pm0.003$ \cr
Haynes et al. (1999) & 1296 & $-7.92\pm0.18$ & $-0.126\pm0.003$ \cr
BBFN (not de-biased) & 467 & $-7.40\pm0.21$ & $-0.135\pm0.004$ \cr
Sakai et al. (2000) & 21 & $-8.07\pm0.72$ & \cr
\noalign{\smallskip}
\noalign{\hrule}
\end{tabular}
}
\end{center}

Can the observed change of slope be caused by systematic errors? 
Any overestimate of $i$ would both lead to an underestimate
of $v_{\rm max}$ and too large a luminosity after correcting for intrinsic
absorption. But to introduce a significant effect on the slope, the errors on
$i$ would have to be large: assuming, e.g., an inclination of 60$\degr$ instead
of 40$\degr$ for an object, the shift in the TFR with a slope of $-6$ would
amount to $-0.93$\,mag.
Such measurement errors are unlikely, as our inclinations do not show any 
dependence on rotation velocity nor apparent scale length nor type.
For that reason, we are also confident that morphologically irregular objects 
in our sample are unlikely to have introduced a significant bias in the 
inclinations.

\section{Discussion}

Our intermediate-$z$ TFR indicates a change in the slope with look--back time.
Most previous studies of the kinematics of spirals in this regime were 
restricted to small data sets ($N\lesssim 20$) and had to assume a constant
slope.
In the ongoing study of \citet{vogt00} of $\sim100$ distant disk galaxies,
it was also assumed that the slope of the local TFR of \citet{PT92} holds
valid at all redshifts. 

CDM-based simulations predict a slight brightening of the $B$-band TFR up to
$z=1$ \citep[e.g.][]{SN99}. In this model, the decrease in mass is more 
than balanced by an increase in luminosity due to younger stellar populations. 
Any mass-dependent evolution of the star formation rate or other TF parameters 
that would cause an evolving TFR slope are yet not implemented in these models.

A change in the TFR slope as we observed makes it impossible to give a single
number for the luminosity evolution of distant galaxies since it depends on the
galaxy mass. The FDF galaxies with $v_{\rm max}\gtrsim150$\,km/s
fall into the same magnitude region as those of the local sample and show no
significant luminosity evolution, whereas the less massive distant
galaxies are brighter by $\sim1-2$ mag.

A mass dependence of the TFR evolution also introduces a dependence on
galaxy type since the various sub-classes of disk galaxies have
different distributions in mass ($\log v_{\rm max}$) for a magnitude--limited 
sample. Thus, a natural explanation of the discrepant results in luminosity 
evolution of previous observational studies is suggested. 
For example, studies based on 
samples with galaxies selected to be blue \citep[like][]{RGCI97} or showing 
strong emission lines \citep[like][]{SP98} are biased towards later types of 
spirals \citep{Vauco61,KK83} resulting in a large luminosity increase. 
On the other hand, studies using mainly galaxies with large disks 
\citep[like][]{VFPGFIK96} may contain a larger fraction of early
and luminous spirals leading to a more modest luminosity evolution.

At last, we speculate on the possible origin of the flatter tilt in all three
redshift bins ($z<0.3, 0.3\le z<0.5, z\ge0.5$). A plausible scenario would be
that we not only trace a pure luminosity evolution but also observe a 
population of blue dwarf galaxies with $M_B>-20$ underrepresented in the local 
Universe. This would be consistent with blue number counts that suggest a 
higher number density of bursting dwarfs at $z\lesssim0.5$ 
\citep[e.g.][]{Ellis97}. 
These objects may have too low a rotation for their luminosity as observed by
\citet{RGCI97}. On the other hand, the star formation rates of all FDF galaxies
($0.2-18 \,M_\sun/$yr) 
fall well below the local regime of starburst galaxies \citep{Kenni98}.
If we exclude the low-mass objects with  
$v_{\rm max}<80$\,km/s and $M_B>-18$ ($N=16$) from the local and distant
sample, we derive slopes of $-7.29\pm0.15$ and $-5.08\pm0.50$, respectively,
confirming the change of tilt with $3\,\sigma$ confidence.

\section{Summary}

We have presented the $B$-band Tully--Fisher relation of
60 disk galaxies at intermediate redshifts. The galaxies comprise a 
magnitude-limited sample within the FORS Deep Field. 
We observe a significant 
($3\,\sigma$) change of slope in comparison to
the local sample.
The larger FDF spirals ($v_{\rm max}\gtrsim150$\,km/s)
fall into a region of the $\log v_{\rm max}$ -- luminosity diagram
consistent with modest or no luminosity evolution. 
On the contrary, the smaller distant galaxies are offset
from the local TFR by order $\sim1-2$\,mag.



\acknowledgments
We acknowledge the continuous support of our project by the PI of the FDF
consortium, Prof. I. Appenzeller. We also thank ESO and the Paranal staff 
for efficient support of the observations.
Our work was funded by the Volkswagen Foundation (I/76\,520), 
the Deutsche Forschungsgemeinschaft (SFB375, SFB439)
and the German Federal Ministry for Education and Science
(ID--Nos. 05\,AV9MGA7, 05\,AV9WM1/2, 05\,AV9VOA).






\clearpage


\end{document}